\documentstyle[12pt,aaspp4]{article}
\slugcomment{Accepted -- To appear in The Astrophysical Journal.}
\def\farcs{\hbox{$.\!\!^{\prime\prime}$}}
\def\fdeg{\hbox{$.\!\!^{\circ}$}}
\lefthead{C.G. Mundell et al.}  

\righthead{Parsec-scale images of flat-spectrum radio sources in Seyferts}

\begin{document}

\title{Parsec-Scale Images of Flat-Spectrum Radio Sources in Seyfert Galaxies}

\author{C.G. Mundell}
\affil{Department of Astronomy, University of Maryland, College Park, MD,
20742, USA;}
\authoremail{cgm@astro.umd.edu}

\author{A.S. Wilson\altaffilmark{1}}
\affil{Department of Astronomy, University of Maryland, College Park, MD,
20742, USA;}
\authoremail{wilson@astro.umd.edu}
\altaffiltext{1}{Adjunct Astronomer, Space Telescope Science Institute}

\author{J.S. Ulvestad}
\affil{National Radio Astronomy Observatory, P.O. Box O, Socorro, NM, 87801,USA;}
\authoremail{julvesta@aoc.nrao.edu}

\author{A.L. Roy\altaffilmark{2}}
\affil{National Radio Astronomy Observatory, P.O. Box O, Socorro, NM, 87801, USA}
\authoremail{aroy@mpifr-bonn.mpg.de}
\altaffiltext{2}{Present Address: MPIfR, Auf dem H\"ugel 69, D-53121 Bonn, Germany}

\begin{abstract}
 
We present high angular resolution ($\sim$2 mas) radio continuum
observations of five Seyfert galaxies with flat-spectrum radio nuclei,
using the VLBA at 8.4 GHz. The goal of the project is to test whether
these flat-spectrum cores represent thermal emission from the
accretion disk, as inferred previously by Gallimore et al. for
NGC~1068, or non-thermal, synchrotron self-absorbed emission, which is
believed to be responsible for more powerful, flat-spectrum nuclear
sources in radio galaxies and quasars. In four sources (T0109$-$383,
NGC~2110, NGC~5252, Mrk~926), the nuclear source is detected but
unresolved by the VLBA, indicating brightness temperatures in excess
of 10$^8$ K and sizes, on average, less than 1 pc. We argue that the
radio emission is non-thermal and synchrotron self-absorbed in these
galaxies, but Doppler boosting by relativistic outflows is not
required. Synchrotron self-absorption brightness temperatures suggest
intrinsic source sizes smaller than $\sim$0.05$-$0.2pc, for these four
galaxies, the smallest of which corresponds to a light-crossing time
of $\sim$60 light days or 10$^4$ gravitational radii for a
10$^8$~M$_\odot$ black hole. In one of these galaxies (NGC~2110),
there is also extended ($\sim$0.2 pc) radio emission along the same
direction as the 400-pc scale jet seen with the VLA, suggesting that
the extended emission comes from the base of the jet. In another
galaxy (NGC~4388), the flat-spectrum nuclear source is undetected by
the VLBA. We also present MERLIN and VLA observations of this galaxy
and argue that the observed, flat-spectrum, nuclear radio emission
represents optically thin, free-free radiation from dense thermal gas
on scales $\simeq$~0.4 to a few pc. It is notable that the two Seyfert
galaxies with detected thermal nuclear radio emission (NGC~1068 and
NGC~4388) both have large X-ray absorbing columns, suggesting that
columns in excess of $\simeq$~10$^{24}$~cm$^{-2}$ are needed for such
disks to be detectable.
\end{abstract}

 
\section{Introduction}

It has become generally accepted that supermassive black holes (SBH)
lie at the center of most, if not all, galaxies (e.g., Richstone et
al., 1998; van der Marel, 1999), with some lying dormant and others
being triggered into an active phase to produce active galactic nuclei
(AGN) (e.g., Haehnelt \& Rees, 1993; Silk \& Rees, 1998). The power
source for this activity is thought to be accretion of material onto
the SBH, with the infalling material forming an accretion disk which,
depending on detailed conditions, then regulates the fueling rate
(e.g. Narayan \& Yi, 1994; Kato, Fukue \& Mineshige, 1998; Blandford
\& Begelman, 1998).  The radius to which these accretion disks extend
(and hence become more easily observable) is not well established, but
current AGN unification schemes advocate a geometrically thick and
clumpy torus (e.g. Krolik \& Begelman, 1988; Krolik and Lepp, 1989;
Pier \& Krolik, 1992) or warped thin disk (Miyoshi et al., 1995;
Greenhill et al., 1995; Herrnstein, Greenhill \& Moran, 1996; Pringle,
1996; Maloney, Begelman \& Pringle, 1996) which hides the nucleus when
viewed edge-on. Our viewing angle with respect to the torus or disk is
then responsible for the observed differences between narrow-line AGNs
(e.g Seyfert 2's), in which our view of the nuclear broad-line region
is obscured (edge-on view), and unobscured (pole-on view) broad-line
AGNs (e.g. Seyfert 1's). Indirect evidence in support of such tori
includes the discovery of broad lines in the polarized (hence
scattered) light of Seyfert 2s (Antonucci \& Miller, 1985; Tran,
1995), sharp-edged bi-cones of ionized gas (e.g., Wilson \& Tsvetanov,
1994) photo-ionized by anisotropic nuclear UV radiation (perhaps
originating from the accretion disk and further collimated by the
torus), large gas column densities (10$^{23-25}$ cm$^{-2}$) to the
nuclei of Seyfert 2's, inferred from photoelectric absorption of soft
X-rays (Turner et al., 1997) and strong mid-infrared emission in both
Seyfert types (e.g., Antonucci, 1993; Alonso-Herrero, Ward \&
Kotilainen, 1996).

Recent, high-resolution studies at optical and radio wavelengths have
begun to provide more direct evidence for `nuclear' disks on
size-scales ranging from the $\sim$100-1000-pc diameter dusty disks
imaged by HST (Jaffe et al., 1993; Ford et al., 1994; Carollo et al.,
1997) and millimeter interferometry (Baker \& Scoville, 1998; Downes
\& Solomon, 1998) to pc-scale disks inferred from HI and free-free
absorption studies (Mundell et al., 1995; Carilli et al., 1998; Peck
\& Taylor, 1998; Wilson et al., 1998; Taylor et al., 1999; Ulvestad,
Wrobel \& Carilli, 1999), down to the 0.25-pc warped, edge-on,
Keplerian maser disk in NGC~4258, imaged by the VLBA (Miyoshi et al.,
1995, Greenhill et al., 1995; Herrnstein, et al., 1996).

Theoretical work indicates that UV/X-ray radiation from the central
engine can heat, ionize and evaporate the gas on the inner edge of the
torus (Pier \& Voit, 1995; Balsara \& Krolik, 1993; Krolik \& Lepp,
1989). Indeed, simple Str\"omgren sphere arguments suggest a radius
for the ionized region of $\rm
~R(pc)~=~1.5~(N_\star/10^{54}~s^{-1})^{1/3}~(n_e/10^5~cm^{-3})^{-2/3}$,
where N$_{\star}$ is the number of nuclear ionizing photons per second
and n$_{\rm e}$ is the electron density. Recalling the typical density
$\rm n_e\sim10^{5-6}~cm^{-3}$ of the ionized disk in NGC~1068 (see
below), we expect R~$\sim$~0.3$-$1.5~pc which is comparable to the
tenths of pc to $\sim$pc-scale resolutions achievable with the VLBA
for nearby Seyferts.  Recent high angular resolution VLBA radio
observations of the archetypal Seyfert 2 galaxy, NGC~1068, by
Gallimore et al. (1997), have shown that emission from one of the
radio components (`S1') may be associated with the inner, ionized edge
of the torus. This radio component has a flat or rising (towards
higher frequencies) spectrum, suggesting it contains the AGN, and a
brightness temperature of up to 4 $\times$ 10$^{6}$~K; it is elongated
perpendicular to the inner radio ejecta and extends over $\sim$40 mas
(3 pc). The radiation mechanism may be free-free thermal emission
(Gallimore et al., 1997), direct synchrotron emission (Roy et al.,
1998) or Thomson scattering of a nuclear flat-spectrum synchrotron
self-absorbed radio core (itself not detected) by the electrons at the
inner edges of the torus (Gallimore et al., 1997; Roy et al., 1998).

This discovery highlights the possibility of using the VLBA to image
the pc-scale disks or tori in other Seyfert galaxies.  However,
flat-spectrum radio sources in AGNs often represent non-thermal
synchrotron self-absorbed radio emission with a much higher brightness
temperature ($>$10$^8$ K) than is characteristic of component S1 in
NGC 1068. High resolution radio observations are thus required to
distinguish between the two emission processes. In the present paper,
we report parsec-scale VLBA imaging of five Seyfert galaxies with
flat-spectrum radio cores and hundred-pc scale, steep-spectrum, radio
jets and lobes. Two of these galaxies also exhibit ionization cones
with sharp, straight edges and axes aligned with the radio ejecta. Our
goal is to determine whether the flat spectrum nuclear radio emission
represents thermal emission from the accretion disk/obscuring torus or
synchrotron self-absorbed emission from a compact radio core source.

The paper is organized as follows; Sections 2 and 3 describe the
sample selection, observations and reduction techniques whilst in
Section 4, the results of the study are presented. Section 5 discusses
possible scenarios for the observed radio emission including direct
non-thermal radiation from the AGN, emission from supernovae or
supernova remnants produced in a starburst, or thermal emission from
the accretion disk. The observed brightness temperatures are discussed
in the context of the NGC 1068 result and comparison is made with
other types of active nuclei such as radio galaxies, radio-loud and
radio-quiet quasars. Section 6 summarizes the conclusions. Throughout,
we assume H$_0$~=~75~km~s$^{-1}$~Mpc$^{-1}$ and q$_0$~=~0.5.
   
\section{Sample Selection}

The radio emission of Seyfert galaxies imaged at resolutions
0\farcs1--1$''$ almost always has the steep spectrum characteristic of
optically thin synchrotron radiation. Flat spectrum cores are rare. In
order to identify galaxies that may contain radio components similar
to `S1' in NGC 1068, we have reviewed both published (e.g., Ulvestad
\& Wilson, 1989, and earlier papers in this series at $\lambda$6 cm and
$\lambda$20 cm; Kukula et al., 1995 at 3.6 cm) and unpublished
(Wilson, Braatz \& Dressel at 3.6 cm) VLA `A' configuration surveys
and other interferometric studies (e.g., Roy et al., 1994). In
selecting candidate galaxies for VLBA observations, we used the
following criteria:

\noindent
$\bullet$ The radio component that is coincident with the optical
nucleus (the position of which is known to $\approx$ 0\farcs2 accuracy
-- e.g., Clements, 1981, 1983), has a flat spectrum ($\alpha$ $\leq$
0.4, S $\propto$ $\nu^{-\alpha}$) between 20 cm and 6 cm or 3.6 cm
with the VLA in `A' configuration. This component must also be
unresolved in the VLA `A' configuration at 2cm and/or 3.6 cm.

\noindent
$\bullet$ The flux density of this component exceeds 5 mJy at 3.6 cm
(for comparison, the total flux density of component `S1' in NGC 1068
at this wavelength is 14 mJy).

\noindent
$\bullet$ There is, in addition, extended, `linear' (double, triple or
jet-like), steep spectrum radio emission on the hundreds of parsecs --
kiloparsec scales, or well-defined, optical ionization cones. The
reason for this last criterion is to define the axis of ejection of
the radio components and thus the expected axis of the accretion disk.

We found only six (excluding NGC 1068) Seyfert galaxies that satisfy
these three criteria in the entire sample of about 130 imaged in the
`A' configuration. We omit one of them because of its unfavorable
declination (--44$^{\circ}$), leaving five for imaging with the
VLBA. These galaxies are T0109$-$383, NGC~2110, NGC~4388, NGC~5252 and
Mrk~926.
 
\section{Observations and Reduction}

\subsection{VLBA Observations}

The observations were obtained with the 10-element VLBA (Napier et al,
1994) at 8.4 GHz during observing runs in 1997 and 1998, details of
which are given in Table 1.  Dual circular polarizations (Right \&
Left) were recorded for all sources, and only the parallel hands
(i.e. RR and LL) were correlated. T0109$-$383, NGC~2110 and Mrk~926
were recorded with a 32-MHz bandwidth and two-bit sampling (8 MHz per
IF, 4 IFs, 2 polarizations) and NGC~4388 and NGC~5252 were recorded
with a 16-MHz bandwidth and two-bit sampling.

The target sources are too weak to obtain estimates of the phase
errors using standard VLBI self-calibration/imaging techniques
(e.g. Walker, 1995); instead the targets were observed in phase
referencing mode, in which frequent observations of a nearby bright
calibrator are interleaved with target scans and used for fringe
fitting, which corrects the large phase errors, delays (phase
variations as a function of frequency) and delay rates (phase
variations as a function of time) present in the data (Beasley \&
Conway, 1995). As described below, extending the coherence time in
this way improves the signal-to-noise ratio and enables an image of
the target source to be made, which can then be used as a starting
model for subsequent cycles of self-calibration.  Target source plus
phase calibrator cycle times are shown in Table 1.  This method is
similar to that used on smaller, connected-element arrays, such as the
VLA (known as `phase calibration'), but is more problematic for VLBI
due to larger and more rapidly varying phase errors.  Rapid changes in
the troposphere at 8.4~GHz therefore require short switching times to
satisfy the condition that the change in atmospheric phase be less
than a radian over the switching interval, thus enabling reliable
phase connection, without 2$\pi$-radian ambiguities, for successful
imaging of the target source (Beasley \& Conway, 1995). In addition,
less frequent observations were made of a bright calibrator (`phase
check') source.

Data editing and calibration followed standard methods (Greisen \&
Murphy, 1998) and used the NRAO Astronomical Image Processing System
({\sc aips}) (van Moorsel, Kemball \& Greisen, 1996). Amplitude scales
were determined from standard VLBA antenna gain tables, maintained by
NRAO staff, and measurements of $T_{\mathrm{sys}}$ made throughout the
run. In addition, all data for source elevations below
$\sim$5$^{\circ}$ were removed, and the antennas at Hancock (HN), and
North Liberty (NL) were deleted from the NGC~5252 dataset as no
fringes were detected to HN, and NL showed poor phase stability due to
bad weather. The final phase corrections, interpolated over time, were
used as a guide for additional data editing.

Despite short switching times between galaxy and phase calibrator,
poor tropospheric conditions and uncertainty in the target source
position prevented immediate imaging of the phase-referenced target
sources using all the data. Observations of the `phase check' source
were therefore used to verify the quality of the phase referencing,
before applying the phase corrections to the target sources, and to
provide ancillary calibration such as manual pulse calibration and
amplitude calibration checks.

After imaging the phase calibrator to verify that the corrections
derived from fringe fitting were valid, phase, delay and rate
corrections were applied to the `check source', from phase calibrator
scans that were adjacent in time to the check source.  Many baselines
displayed poor phase coherence at some point in the observing run,
preventing a coherent image of the source from being produced
initially from the whole dataset. Instead, small time ranges
(e.g. around 1 hour), within which the majority of antennas had less
rapidly varying phases, were selected to be used in the initial stages
of the imaging process. The `check' source, with calibration applied
from the phase calibrator, was imaged for the selected small time
range. The resultant image was then used as an input/starting model
for subsequent cycles of self-calibration. This self-calibration
process then enabled the remaining data to be fully calibrated and
used to make a final image of the `check' source. The final structure,
flux and position of each `check' source compared well with previously
published images (e.g. Browne et al., 1998; Fey \& Charlot, 1997) and
images produced from our data using self-calibration alone.  This
method provides an independent consistency check on the phase
referencing, increasing our confidence in the images of the target
sources. Only one `check' source (J0044-3530) was not successfully
imaged due to insufficient data (i.e. only 3 minutes at very low
elevation). The target sources were then imaged using the same method,
with natural and uniform data weighting. The uniformly weighted images
(with robust parameter 0 - Briggs, 1995) are shown in Figure 1. The
naturally weighted images, with more sensitivity to extended emission,
were used to derive the brightness temperature limits to possible
thermal emission from the program galaxies; these limits are
$\sim$30\% lower than those derived from the uniformly weighted
images shown in this paper.

The uncertainty in the flux  scale is taken to be $\sim$5\% and is
included in the total uncertainties in flux densities quoted in Table
2. These errors were derived by adding, in quadrature, the 5\%
amplitude scale error, the r.m.s. noise in the final image and the error
in the Gaussian fitting.

The accuracy of the target source positions is dominated by the
uncertainty in the position of the phase calibrators ($\sim$0.4 -- 14
mas; see Table 1). Additional positional errors, due to the transfer of
phase corrections from the phase calibrator to the target source, are
negligible due to the promixity of each calibrator to its target
source.

\subsection{MERLIN observations}

NGC~4388 was not detected by the present VLBA observations. We
therefore obtained and analyzed MERLIN $\lambda$6-cm (4.993-GHz)
archival data for NGC~4388, which was observed on 7th December, 1992
with six antennas. Phase referencing was performed with regular
observations of 1215+113, interleaved throughout the observing run and
3C286 was used for flux and bandpass calibration. A flux of 7.087 Jy
for 3C286 was adopted, assuming a total flux density of 7.382 Jy
(Baars et al., 1977) and correcting for MERLIN resolution
effects. After initial gain-elevation corrections and amplitude
calibration using MERLIN software, the data were transferred to {\sc
aips} for all subsequent phase and amplitude calibration, data editing
and imaging. Dual polarizations were recorded for a 15-MHz bandwidth,
centered at 4.993 GHz, but the right circular polarization data were
removed due to instrumental problems, resulting in a final image of
the left circular polarization only (Figure 2).

\section{Results}

Five flat-spectrum-core Seyferts, were observed with the VLBA at 8.4
GHz. Four of the five sources were detected (T0109$-$383, NGC~2110, NGC
5252, Mrk~926) and show compact, unresolved cores with brightness
temperatures T$_{\rm B}$ $>$10$^8$ K, total luminosities at 8.4 GHz
of $\sim$10$^{21}$ W Hz$^{-1}$ and sizes, on average, less than 1
pc. In addition to the core emission, NGC~2110 shows extended emission
which may represent the inner parts of the radio jets, and NGC~5252
may be marginally extended (Figure 1). NGC~4388 is not detected with
the VLBA, but is detected at 5 GHz with MERLIN (Figure 2).  We find no
evidence for emission (to a 3-$\sigma$ limit of T$_{\rm B}$
$\sim$10$^6$ K) extended perpendicular to the hundred-pc scale radio
emission in T0109$-$383, NGC~2110, NGC~5252 or Mrk~926, as would be
expected for emission from an accretion disk, but we discuss the
possibility of thermal emission from NGC~4388 (Section 5.4). The
measured and derived properties of each source are listed in Table 2,
while more detailed properties of NGC~2110 and NGC~4388 are given in
Tables 3 and 4 respectively. The properties of each source are
discussed more fully below. Distances are calculated assuming
H$_0$~=~75~km~s$^{-1}$~Mpc$^{-1}$ and q$_0$~=~0.5, except for NGC~4388
which is assumed to be at the distance of the Virgo cluster, taken to
be 16 Mpc.

\subsection{T0109$-$383}

T0109$-$383 (NGC~424) is a highly inclined ($\sim$75$^{\circ}$)
early-type ((R)SB(r)0/a -- de Vaucouleurs et al. 1991) Seyfert galaxy
at a distance of 46.6 Mpc. The nucleus of T0109$-$383, originally
classified as a Seyfert type 2 (Smith, 1975), exhibits strong line
emission from highly ionized species such as [Fe {\sc
vii}]$\lambda$5720,6086 and [Fe {\sc x}]$\lambda$6374 (Fosbury \&
Sansom, 1983; Penston et al., 1984). Analysis of the continuum
emission from the far IR to the far UV and decomposition of the
H$\alpha$ -- [N{\sc ii}] blend led Boisson \& Durret (1986) to suggest
a re-classification of T0109$-$383 to a Seyfert 1. The recent
discovery of broad components to the H$\alpha$ and H$\beta$ lines,
along with emission from Fe {\sc ii}, confirms the type 1
classification (Murayama et al., 1998). VLA images of the radio
emission at 6 and 20 cm, show the nuclear radio source to consist of
an unresolved core with a flat spectrum
($\alpha_6^{20}$=0.17$\pm$0.07) between $\lambda$6 and $\lambda$20 cm,
and a weaker, secondary, steep spectrum component $\simeq$1\farcs4
east of the core (Ulvestad \& Wilson, 1989). Similar radio structure
is seen in the 8.4-GHz VLA image (Braatz, Wilson \& Dressel,
unpublished), shown in Figure 1, with the core spectrum remaining
relatively flat ($\alpha^6_{3.5}$=0.21) between 6 and 3.5 cm (Morganti
et al., 1999). The results of Gaussian fitting to the 8.4-GHz VLBA
image (Figure 1), given in Table 2, show the sub-pc scale nuclear
emission to be unresolved, with a peak brightness of T$_{\rm
B}$~$>$~8.1~$\times$~10$^8$~K, adopting a source size smaller than
half of the beamsize. The peak and integrated 8.4-GHz VLA fluxes for
the core, 10.4 mJy beam$^{-1}$ and 11.2 mJy respectively, are in
excellent agreement with those measured from the VLBA image (Table 2),
indicating that little nuclear emission was missed by the VLBA.  A
similar peak brightness of 10.4~mJy beam$^{-1}$ is found in the 3.5-cm
ATCA image of Morganti et al. (1999), while their slightly higher
integrated flux includes some of the emission $\simeq$ 1$''$ E and W
of the nucleus (Ulvestad \& Wilson, 1989; Figure 1). The excellent
agreement between the nuclear $\lambda$3.6-cm fluxes in observations
spanning $\sim$six years indicates no significant variability.

In the VLBA image, we detected no extended emission in the N-S
direction (as might be expected from a parsec-scale, thermal disk if
the arcsec-scale, steep spectrum, E-W emission in the VLA image is
interpreted as emission from nuclear ejecta) brighter than
$\sim$1.3~$\times$~10$^6$~K (3$\sigma$ in the naturally weighted
image) and more extended than 0.27 pc (half of the beamsize in the
naturally weighted image).

\subsection{NGC~2110}

NGC~2110 was initially classified as a Narrow Line X-ray Galaxy, NLXG,
(Bradt et al., 1978), and lies in an S0/E host galaxy (Wilson, Baldwin
\& Ulvestad, 1985) at a distance of 30.4 Mpc. Such NLXG's have a
sufficient column of dust to the nucleus to obscure the broad line
region, thus leading to a Seyfert 2 classification of the optical
spectrum, but an insufficient gas column to attenuate the 2--10 keV
emission, so the hard X-ray luminosity is comparable to those of
Seyfert 1's (Weaver et al., 1995a; Malaguti et al., 1999). Early radio
observations found NGC~2110 to be a strong radio source (Bradt et al.,
1978) and subsequent VLA imaging (Ulvestad \& Wilson, 1983; 1984b)
showed symmetrical, jet-like radio emission, extending $\sim$4$''$ in
the N-S direction and straddling a central compact core.  A more
recent VLA A-configuration image at $\lambda$3.6 cm, obtained by Nagar
et al. (1999) and shown in Figure 1, contains a wealth of complex
structure. Ulvestad \& Wilson (1983) found the spectrum of the core to
be relatively flat (spectral index $\alpha_6^{20}$
$\sim$0.36$\pm$0.05) between $\lambda$20 cm and $\lambda$6 cm, but
becoming steeper ($\alpha_2^{6}$$\sim$0.96$\pm$0.09) between
$\lambda$6 cm and $\lambda$2 cm (assuming no time variability). Using
the $\lambda$3.6-cm core flux measurement of Nagar et al. (1999) and
ignoring variability or resolution effects gives spectral indices of
$\alpha_{3.6}^6$ = 0.61 and $\alpha_2^{3.6}$ = 1.31, also suggesting a
steepening of the spectrum at higher frequencies.

The radio continuum emission of NGC~2110, imaged with the VLBA at
$\lambda$3.6 cm and shown in Figure 1, consists of a compact core,
presumably the nucleus, and slightly extended emission which is most
pronounced to the north. The results of fitting a single-component
Gaussian are given in Table 2; the fact that the integrated flux is
significantly higher than the peak flux also suggests the source is
resolved. Resolved structure is also evident in the time-averaged
$(u,v)$ data (not shown), consistent with an unresolved point source
(with a flux density of $\sim$ 8 mJy) superimposed on an extended
``halo'' with approximate dimensions of 2.5 (N-S) $\times$ 0.5
(E-W)~mas.  Preliminary two-component Gaussian fits to the image are
also consistent with an unresolved point source and an extended
component. We therefore subtracted an 8-mJy point source (in the
$(u,v)$ plane using the {\sc aips} task {\sc uvsub}) positioned at the
peak of the 3.6 cm VLBA image, and studied the residual emission. This
emission is extended both north and south of the core by $\sim$0.7
mas, consistent with emission from the inner regions of the northern
and southern jets.

Using the brightness of 8 mJy beam$^{-1}$ for the unresolved component
and assuming an upper limit to the source size of 0.94 $\times$ 0.36
mas (half of the beamsize), we find T$_{\rm B}$ $>$ 6.0 $\times$
10$^8$~K.  In addition to the core and extended emission, the Gaussian
fits suggest the presence of a third component, centered $\sim$1.95 mas
north of the core; its size and direction of elongation are not well
constrained. This component may be a knot in the northern jet. A
summary of the fitted properties of each component is given in Table
3.

The total VLBA-detected flux density of the source (zero baseline flux
measured in the {\sc uv} plane) is 30 mJy.  This flux density is lower
than the previously measured VLA core flux of 77.6 mJy at this frequency
(Nagar et al., 1999), presumably due to the high spatial resolution of
the present observations and missing short spacings of the VLBA
compared to the VLA, thereby reducing our sensitivity to extended
structure. This may also explain why we detect no VLBA counterpart to
the small eastern extension present in the $\lambda$3.6-cm VLA image,
which contains about 3.6 mJy of flux and extends approximately
0\farcs5 east of the core (Nagar et al., 1999). Alternatively, the
extension in the VLA image may be a result of instrumental effects
caused by the source position being close to the celestial equator and the
short duration of the snapshot observation, an effect termed `equator
disease' (Antonucci \& Ulvestad, 1985).  In the VLBA image, we detect no
extended emission in the E-W direction (such as might be expected from a
parsec-scale thermal disk) brighter than 3.1 $\times$
10$^6$ K (3$\sigma$ in the naturally weighted image), and more extended
than 0.07 pc (one half of the E-W beamsize in the naturally weighted
image).

\subsection{NGC~4388}

NGC~4388 is a nearby, edge-on spiral galaxy (SB(s)b pec - Phillips \&
Malin, 1982) which is thought to lie close to the centre of the Virgo
cluster (Phillips \& Malin, 1982) and may be tidally disturbed by
nearby cluster core galaxies M84 or IC3303 (Corbin, Baldwin \& Wilson,
1988). Ionization cones extend approximately perpendicular to the disk
(Pogge, 1988; Corbin et al., 1988; Falcke, Wilson \& Simpson, 1998)
and the kinematics of the ionized gas in the narrow line region (NLR)
shows a complex combination of rotation and outflow (Corbin et
al. 1988; Veilleux, 1991; Veilleux et al., 1999). The nucleus is
variously classified as Seyfert type 1 or 2, with the high galactic
inclination and obscuring dust lanes making unambiguous classification
difficult (Falcke et al., 1998). Shields \& Filippenko (1988) report
broad, off-nuclear H$\alpha$ emission, but subsequent IR searches for
broad lines such as Pa$\beta$ (Blanco, Ward \& Wright, 1990; Ruiz,
Rieke \& Schmidt, 1994) and Br$\alpha$ and Br$\gamma$ (Veilleux,
Goodrich \& Hill, 1997) have failed to detect a broad nuclear
component.
 
Previous radio continuum images of NGC~4388 (Stone et al., 1988;
Carral, Turner \& Ho, 1990; Hummel \& Saikia, 1991; Falcke et al.,
1998) show complex, extended structure, both along the galactic plane
and perpendicular to it. A recent 3.5 cm VLA image of the extended radio
emission (Falcke et al., 1998) shows, in more detail, the
`hour-glass'-shaped radio outflow to the north of the galactic plane,
and the compact ($\sim$1\farcs9 separation) central double, which were
suggested by earlier images. In Section 4.3.1 we concentrate on the
radio emission from the northern component of the compact radio
double, which shows a flat spectrum up to 2 cm (Carral, Turner \& Ho,
1990) and is thought to be the nucleus, and in Section 4.3.2, we
discuss the extended emission to the SW.

\subsubsection{The nucleus}

As stated earlier, NGC~4388 is not detected in the 8.4-GHz VLBA
observations, with a 3-$\sigma$ brightness temperature limit of
T$_{\rm B}$ $\lesssim$2.2~$\times$~10$^6$ K ($\sigma$=63.2
$\mu$Jy/beam with a beam size of 2.52~$\times$~1.46 mas in the
naturally weighted map, with a factor 1.7 applied to correct for
decorrelation due to residual imperfections in the phase referencing
corrections, estimated using the check source). We do, however, detect
emission from NGC~4388 at $\lambda$6cm with MERLIN. The uniformly
weighted MERLIN image (Figure 2) shows emission from two components,
labelled M1 and M2, the stronger of which we identify with the nucleus
and discuss in more detail here, while M2 is discussed in Section
4.3.2.  The nuclear component M1, has a peak brightness of 1.2 mJy
beam$^{-1}$ which corresponds to a brightness temperature T$_{\rm B}$
$>$ 2.4 $\times$ 10$^4$ K at 5 GHz (beamsize 91 $\times$ 39.5 mas, see
Table 4).  The nucleus is unresolved in the MERLIN data, indicating
that the source size is intermediate between the MERLIN and VLBA beam
sizes.  However, a combination of the MERLIN and VLBA results with
published spectral index information can further constrain the source
size.

Earlier radio observations of NGC~4388 have found the nuclear spectrum
to be flat from 1.49 GHz to 15 GHz. The spectral index was measured to
be $\alpha$ = 0.26 between 1.49 GHz and 4.86 GHz with a relatively
large beamsize of 1\farcs2 (Hummel \& Saikia, 1991) and Carral et al.
(1990) report a flat spectrum up to 15 GHz with an upper limit to the
nuclear size of 70 mas. Including the VLA 8.4 GHz core flux of Kukula
et al (1995) suggests that the spectrum of the nucleus may be very
slightly inverted between 8.4 GHz and 15 GHz ($\alpha$ = --0.05) but
within the errors it can be taken as flat. We therefore used the
measured MERLIN 5-GHz peak flux to derive {\em predicted} VLBA 8.4 GHz
fluxes of the nucleus, for spectral indices of both $\alpha$ = 0.0 and
0.26, and converted these predicted fluxes to brightness temperatures,
assuming the source is unresolved by the representative VLBA beamsize
of 2.52 $\times$ 1.46 mas.

These predicted temperatures are listed in Table 4 and are above the
detection threshold of the VLBA observations for a source size equal
to or smaller than the VLBA beam. The larger predicted brightness
temperature, for a source size equal to the VLBA beam, of T$_{\rm B}$
$\simeq$ 8.3 $\times$ 10$^6$ K is, however, only 3.8 times greater
than our 3-$\sigma$, VLBA detection limit and so the solid angle of the
source need only be 3.8 times larger than the VLBA beamsize to be
undetected. We therefore constrain the size of the nucleus to be
$\gtrsim$ 3.7 mas ($\alpha$=0.0) or $\gtrsim$ 0.3 pc.  Sensitive, high
angular resolution VLBA observations at lower frequencies such as 2.3
GHz and 1.4 GHz are required to determine the actual size and
structure of the nucleus in NGC~4388.

\subsubsection{Collimated radio emission}

In addition to the core emission at 5 GHz, the MERLIN image of
NGC~4388 shows a second weak component SW of the nucleus (labelled M2
in Figure 2). This component is only $\sim$0.25$''$ away from the
nucleus, lying along PA $\sim$211$^{\circ}$ with respect to the core,
and should not be confused with the stronger, more distant radio
component seen in previous VLA images (e.g. Falcke et al, 1998),
which lies $\sim$1\farcs9 from the nucleus (in PA
$\sim$201$^{\circ}$). To establish the reality of this weak
MERLIN component, the VLA 8.4-GHz data (published by Falcke et al.,
1998) were re-examined. The uniformly weighted image, shown in Figure
3a, has a rather elongated synthesized beam of size 0\farcs7 $\times$
0\farcs22 (PA 74$^{\circ}$), but a bridge of emission is clearly
visible, connecting the two main radio components (labelled V1 \&
V2). A similar bridge of emission was seen in the 4.86-GHz VLA image
of Hummel \& Saikia (1991) and the 8.4-GHz VLA snapshot image of
Kukula et al. (1995). The crosses indicate the positions of the
nucleus and the weak component (M2) visible in the MERLIN 5 GHz
image. Figure 3b shows a super-resolved image produced from the same
VLA data, with a circular synthesized beam of 0\farcs22. Again the
crosses mark the MERLIN components, and an extension of the 8.4 GHz
VLA emission is seen at the location of the weak MERLIN component M2,
further suggesting that the MERLIN emission is real. The
super-resolved image also suggests that a well-collimated jet of
emission is emanating from the core (V1) along PA $\sim$210${^\circ}$
for $\sim$1\farcs5, before changing direction at V2. The MERLIN
component M2 would then be an inner part of this collimated radio
ejection. Falcke et al (1998) find a good association between V2 and
a `spike' of optical line emission, suggesting interaction of the
radio jet with a cloud in the NLR. The `hooked' shape of the southern
radio jet, suggested by Fig. 3b, is reminiscent of similarly
well-collimated radio jets seen in an increasing number of Seyferts
(e.g., Mrk~34 - Falcke et al., 1998; Mrk~3 - Kukula et al. 1993;
NGC~1068 - Wilson \& Ulvestad, 1983; NGC~2110 - Ulvestad \& Wilson,
1983; Nagar et al., 1999), which are deflected or terminate at radio
hot spots. However, sensitive high angular resolution radio
observations are required to image the detailed structure of the
proposed radio jet in NGC~4388.

\subsection{NGC~5252}

NGC~5252, an S0 galaxy (de Vaucouleurs et al., 1991) with a
Seyfert type 1.9 nucleus (Acosta-Pulido et al., 1996), exhibits a
dramatic bi-cone of ionized gas (Tadhunter \& Tsvetanov, 1989; Wilson
\& Tsvetanov, 1994; Acosta-Pulido et al., 1996) extending $\sim$35
kpc from the nucleus (along PA 165$^{\circ}$) and argued to be ionized
by anisotropic nuclear UV radiation. The intrinsic anisotropy of the
ionizing radiation was nicely confirmed by HI $\lambda$21-cm images,
which show neutral hydrogen filling the regions outside the bi-cone
(Prieto \& Freudling, 1993, 1996).

Observations with the VLA (Wilson \& Tsvetanov, 1994) show a radio
structure consisting of a central, compact core, with a relatively
flat spectrum between $\lambda$20 cm and $\lambda$6 cm
($\alpha^{20}_6$ = 0.22), steepening at shorter wavelengths
($\alpha^6_{3.6}$ = 0.78), and weaker emission extending $\sim$2$''$
north (PA $\sim$345$^{\circ}$) and south (PA $\sim$175$^{\circ}$) of
the core. A second compact radio component, seen $\sim$22$''$ north
(PA --8\fdeg9) of the nucleus, may be associated with NGC~5252,
lying close to the PA of the extended core emission and coinciding
with a region of high excitation ionized gas (Wilson \& Tsvetanov,
1994), or may be a background source (Morse et al., 1998).  More
recent radio images by Nagar et al. (1999) at $\lambda$20 cm and
$\lambda$3.6 cm confirm the overall spectrum of the core
($\alpha^{20}_{3.6}$ = 0.32) and the radio continuum features seen in
the earlier images.

The 8.4-GHz VLBA image of the core of NGC~5252 is shown in Figure 1,
along with the larger scale $\lambda$20-cm VLA image (from Nagar et
al., 1999), and the results of Gaussian fitting to the emission are
detailed in Table 2.  The deconvolved size of the core is less than
half of the beamsize and so we classify the emission as
unresolved. The peak brightness of 7.9 mJy beam$^{-1}$ (corresponding
to T$_{\rm B}$ $>$ 4.2 $\times$ 10$^8$ K) is in agreement with that
measured by Nagar et al. (1999) -- 7.9 mJy beam$^{-1}$ -- and slightly
higher than that found by Wilson \& Tsvetanov (1994) -- 6.7 mJy
beam$^{-1}$.  The integrated VLBA flux (9.1 mJy) is also very similar
to that of Nagar et al. (1999) (9.3 mJy).  The similarity of the VLBA-
and VLA-measured nuclear fluxes indicates that little emission was
missed by the VLBA.  The fact that the measured integrated VLBA flux
is $\sim$15\% higher than the peak and the possible visible extension
to the SW (Fig. 1) are evidence for weak extended emission.  The
quality of the data, however, make this detection tentative and higher
sensitivity observations are required to confirm the extension and
determine its structure. 

\subsection{Mrk~926}

Mrk~926 (MCG-2-58-22), the most distant object in our sample at
$z$=0.0473, is a type 1 Seyfert and was first identified as a Seyfert
through its luminous (L$_x$ = 5 $\times$ 10$^{44}$ erg~s$^{-1}$)
X-ray emission (Ward et al. 1978). Later X-ray observations found a
relatively flat X-ray spectrum (although detailed model fits are
controversial - George et al., 1998) and Weaver et al. (1995b)
suggested that their ASCA observations of Mrk~926 were consistent with
a ``bare'' Seyfert 1 nucleus.  X ray variability, with a 14 year
timescale, provides an upper limit of 4.3 pc for the size of the
emitting region (Weaver et al., 1995b).

The 8.4 GHz VLA image (Braatz, Wilson \& Dressel, unpublished), shown
in Figure 1, shows a compact core with weak E-W extensions confirming
the earlier 6-cm image (Ulvestad \& Wilson, 1984a). The nuclear radio
spectrum is relatively flat between $\lambda$6 cm and $\lambda$3.6
($\alpha^{6}_{3.6}$ = 0.24), but steepens at longer wavelengths
($\alpha^{20}_{6}$ = 1.1). This steepening of the spectrum may, in
part, be due to inclusion of the extended E-W emission in the low
resolution $\lambda$20-cm observation (Wilson \& Meurs, 1982), but is
also consistent with a lack of free-free absorption toward the
nucleus, supporting the X-ray classification as a ``bare'' Seyfert 1
nucleus.  The 8.4-GHz VLBA image (Figure 1) shows an unresolved
source. The results of Gaussian fitting to the VLBA detection of
Mrk~926 and corresponding derived quantities are given in Table 2.  A
peak brightness of 7.5 mJy beam $^{-1}$ is measured in the VLA image
while the peak brightness in the VLBA image is 4.6 mJy beam$^{-1}$. No
additional extended flux is present in the VLBA data as can be seen
from the image (to a 3-$\sigma$ limit of T$_{\rm B}$ $<$ 1.3 $\times$
10$^6$ K) and the agreement between the peak and integrated flux
densities (Table 2). The upper limit to the size of the unresolved
core is $\sim$1.3 $\times$ 0.5 pc (half of the beamsize).  The fact
that less flux is detected with the VLBA than the VLA, however,
suggests additional emission, extended on scales between the VLA
(0\farcs27 $\times$ 0\farcs21) and VLBA beams (2.8 $\times$ 1.0 mas),
or time variability.

\section{Discussion}

\subsection{Compact Cores and Flat Spectra}

In contrast to the commonly observed high brightness temperature, flat
spectrum cores in quasars, blazars and radio galaxies (e.g. Pearson et
al., 1998; Kellerman et al., 1998, Hough et al., 1999), flat spectrum
cores in Seyfert galaxies appear to be rare (e.g. de Bruyn \& Wilson
1978; Sadler et al., 1995; Morganti et al., 1999), with only
$\sim$10\% detected in the sample of de Bruyn and Wilson (1978).  High
brightness temperature, flat spectrum cores are thought to represent
the synchrotron self-absorbed base of a relativistic jet produced by
the central engine (e.g., Peterson, 1997), while the steep spectrum
components are considered to be associated with shocks along the
jet. Despite the small linear extents of Seyfert radio jets compared
with those in quasars and radio galaxies, the relative proximity of
Seyferts permits higher linear resolution to be achieved at comparable
angular resolutions to studies of quasars and radio galaxies. However,
various factors, such as free-free absorption by broad line region
(BLR) or other dense gas and dominance by steep spectrum emission, may
make it difficult to detect flat spectrum radio cores in Seyfert
galaxies.  In addition, evidence is mounting that Seyfert radio jets
are not relativistic (Ulvestad et al., 1999), but instead may be
thermally dominated (Bicknell et al., 1998; Wilson \& Raymond, 1999),
thereby removing Doppler boosting of the putative flat spectrum core
emission as a means for increasing its detectability.

As shown by recent studies of NGC~1068 (Gallimore et al., 1997; Roy et
al., 1998) an alternative explanation for flat spectrum cores in
Seyferts is emission from the obscuring torus itself, either as
thermal bremsstrahlung (Gallimore et al., 1997), direct synchrotron
radiation (Roy et al., 1998) or scattered synchrotron radiation from a
self-absorbed core (Gallimore et al., 1997; Roy et al., 1998). Here,
we discuss our results in this context, examining whether the emission
is likely to be thermal or non-thermal, central engine or starburst
related and whether, if associated with a central engine, Doppler
boosting could be important.

\subsection{Nuclear emission - thermal or non-thermal?}

High angular resolution 8.4-GHz VLBA observations of the flat-spectrum
nuclear component, `S1', in NGC~1068 have revealed an edge-on,
disk-like structure, aligned roughly perpendicular to the radio jet
axis, with an extent of $\sim$0.5 pc and a brightness temperature
ranging from $\sim$5 $\times$ 10$^5$ to 3.7 $\times$ 10$^6$ K
(Gallimore et al., 1997). In contrast, we find high T$_{\rm B}$
($>$10$^8$ K), unresolved ($<$1 pc) cores in T0109$-$383, NGC~2110,
NGC~5252 and Mrk~926, consistent with non-thermal emission. The
brightness temperature of a synchrotron self-absorbed source is
T$_{\rm B}$~$\simeq$~$\gamma$m$_{\rm
e}$c$^2$/3k~$\simeq$~2.0~$\times$~10$^9$$\gamma$ K, where $\gamma$ is
the Lorentz factor, m$_{\rm e}$ the electron mass, c the speed of
light and k Boltzmann's constant. The maximum T$_{\rm B}$ for such a
source is $\sim$10$^{12}$ K, limited by the ``inverse Compton
catastrophe'', in which cooling by inverse Compton scattering quickly
reduces the brightness temperature to 10$^{11-12}$ K (e.g., Kellermann
\& Pauliny-Toth, 1969; Kellermann \& Pauliny-Toth, 1981).  Readhead et
al. (1994) argue that a more physical limit is the equipartition
brightness temperature, which limits the intrinsic brightness
temperature in the emission rest frame to be $\sim$10$^{11}$ K
(consistent with average observed values - L\"ahteenm\"aki, Valtaoja
\& Wiik, 1999). Although observed T$_{\rm B}$'s as high as 10$^{16}$ K
have been inferred from intra-day variability seen in some blazars
(e.g., Crusius-Waetzel \& Lesch, 1998), and in one extreme case,
intra-hour variability in the quasar PKS~0405$-$385, suggesting
T$_{\rm B}$~$\sim$~10$^{21}$~K if the variability is intrinsic to the
source or T$_{\rm B}$~$>$~5~$\times$~10$^{14}$~K if explained by
interstellar scintillation of a source smaller than 5$\mu$as
(Kedziora-Chudczer et al., 1997), the brightness of these objects is
thought to be extremely Doppler boosted by relativistic outflows.  Our
measured lower limits to the T$_{\rm B}$ of T0109$-$383, NGC~2110,
NGC~5252 and Mrk~926 are in the range (2$-$8)~$\times$~10$^8$~K and
are therefore consistent with synchrotron self-absorption, but do not
require Doppler boosting. For $\gamma$=1, the intrinsic brightness
temperature of a self-absorbed source is 2~$\times$~10$^9$~K,
suggesting source sizes smaller than $\sim$0.2$-$0.3 mas, or
$\sim$0.05$-$0.2pc, for these four galaxies, the smallest of which
corresponds to a light-crossing time of $\sim$60 light days or 10$^4$
gravitational radii for a 10$^8$~M$_\odot$ black hole. Such small
sources may be resolved at 8 GHz with future space VLBI missions such
as ARISE (Ulvestad, Gurvits \& Linfield, 1997; Ulvestad \& Linfield,
1998).

The spectra of the nuclei of NGC~5252 and NGC~2110 appear to steepen
towards higher frequencies (as discussed in Section 4), consistent
with the cores becoming optically thin to synchrotron self-absorption
at $\sim$$\lambda$6 cm (as suggested for NGC~5252 by Wilson \&
Tsvetanov, 1994). Alternatively, free-free absorption by ionized gas,
with optical depths $\tau_{\rm ff}$(1.4 GHz)~$\lesssim$~0.6, could
account for the observed spectral indices between $\lambda$20~cm and
$\lambda$6~cm for T0109, NGC2110 and NGC5252, assuming that the
intrinsic emission is optically thin synchrotron with
$\alpha^{20}_6$~=~0.7.

No evidence of thermal disk-like emission, extended perpendicular to
the collimation axis, is found in these sources to a 3-$\sigma$ limit
of $\sim$10$^6$ K (see Figure 1). Any emission similar to the
brighter, inner (0.5-pc) disk in NGC~1068, could just be spatially
resolved in T0109$-$383, NGC~2110 and NGC~5252, but not in more
distant objects like Mrk~926. Nevertheless, these four Seyferts are
dominated by the compact, high T$_{\rm B}$ core emission which is
completely different from NGC~1068, which shows no compact, unresolved
core.  The absence of a bright core in NGC~1068 may reflect absorption
by a BLR cloud or by the (probably) thermal disk seen in radio
continuum. As we discuss below, a high column density of ionized gas
in such a disk is needed for detectable thermal radio emission and the
disk could hide the compact, high brightness core through free-free
absorption.  The column density towards the nucleus of NGC~1068 is
thought to be so high ($>$10$^{26}$ cm$^{-2}$ -- Matt et al., 1997)
that it is optically thick to Compton scattering and the nucleus is
totally hidden from view even in hard X-rays. The X-ray inferred
column densities for three of the core-dominated Seyferts are
significantly lower at 2.4 $\times$ 10$^{22}$ cm$^{-2}$ for NGC~2110
(Weaver et al., 1995a), 3.4 $\times$ 10$^{22}$ cm$^{-2}$ for NGC~5252
(Turner et al., 1997) and $\sim$5.7 $\times$ 10$^{20}$ cm$^{-2}$ for
Mrk~926 (Weaver et al., 1995b).  

Given the detection of the sub-pc scale, non-thermal radio sources,
the disk cannot be optically thick to free-free absorption if the
synchrotron self-absorbed radio core is seen through it. We may then
calculate an upper limit to the thermal radio emission from the disk
by assuming $\tau_{\rm ff}$(8.4 GHz)~$\lesssim$~0.5 and that the
entire column density inferred from X-ray photoelectric absorption is
fully ionized.  As the disks in NGC~2110 and NGC~5252 are expected to
lie along the beam minor axis, we assume a conservative upper limit to
the disk diameter of twice the minor axis beamsize and use the height
to diameter ratio of 2:1 as found for NGC~1068, giving maximum disk
dimensions of 0.22~$\times$~0.11~pc and 0.86~$\times$~0.43~pc for
NGC~2110 and NGC~5252 respectively\footnote{\small An edge-on disk
geometry is unlikely for Mrk~926, a Seyfert 1, and the inferred column
density for this galaxy is much lower than that of NGC~2110 or
NGC~5252. Thus the predicted radio flux from Mrk~926 will be lower
than from NGC~2110 or NGC~5252.}. The assumed upper limits to the disk
size, column density and free-free optical depth yield lower limits to
both the electron densities n$_{\rm
e}$~$\sim$~7.3~$\times$~10$^{4}$~cm$^{-3}$ and
2.6~$\times$~10$^{4}$~cm$^{-3}$, and electron temperatures T$_{\rm
e}$~$\sim$~3~$\times$~10$^4$~K and 1.8$\times$~10$^4$~K for NGC~2110
and NGC~5252 respectively, and therefore a maximum predicted 8.4-GHz
flux of 1.4 $\mu$Jy for both, undetectable with present
observations. These calculations indicate that thermal radio emission
from an accretion disk is most likely to be detected from nuclei with
high ($>$10$^{24}$ cm$^{-2}$) X-ray inferred column densities, as may
be the case for NGC~4388 (see Section 5.4). Thus, further insight into
the nature of thermal radio cores in Seyferts might be gained with
milliarcsecond-resolution observations of galaxies with high
($>$10$^{24}$ cm$^{-2}$) X-ray inferred column densities.

\subsection{Seyfert nuclei - starburst or accretion-powered central engine?}

A key question is whether the nuclear power source in radio-quiet
quasars, Seyferts and ultra-luminous infrared galaxies is a compact
starburst or accretion onto a supermassive black hole.  Whilst
emission from hot stars in the torus might account for the featureless
continua in Seyfert 2's (Fernandes \& Terlevich 1995;
Gonz\'alez-Delgado et al. 1998), it seems that starbursts cannot
provide the necessary collimation to produce radio jets. The existence
of radio jets is, therefore, often used as an indication of the
presence of a black hole plus accretion disk. Although some Seyferts
are now known to possess strikingly collimated jets (e.g., Nagar et
al., 1999; Kukula et al., 1999), the resolution of the radio images is
often insufficient to demonstrate the high degree of collimation seen
in radio galaxies and radio-loud quasars.

Although compact starbursts may co-exist with AGNs in some Seyferts
(e.g., Heckman et al., 1997; Gonz\'alez-Delgado et al., 1998; Carilli
et al., 1998), we have argued that the sub-pc scale, high brightness
radio emission from T0109$-$383, NGC~2110, NGC~5252 and Mrk~926 is
dominated by the central engine. Radio emission from a starburst
region consists of synchrotron radiation from supernova remnants
(SNR's) plus thermal free-free emission from HII regions. The
brightness temperature of such a region cannot exceed 10$^5$ K at
$\nu>$ 1 GHz (Condon, 1992) and so the high brightnesses of our
Seyfert nuclei, along with their sub-pc sizes rule out a starburst
origin for the radio emission.  In fact, these Seyfert nuclei show
similar brightness temperatures to some radio-quiet quasars (Blundell
\& Beasley, 1998) and LINERs (Falcke et al., 1999), which are also
argued to be non-thermal emission from black-hole powered central
engines rather than compact starbursts.  In NGC~4388, the
well-collimated radio jet also suggests the presence of an AGN,
despite the lack of an unresolved, high brightness nucleus.

We can also rule out individual or a collection of extremely bright
radio supernovae (RSN) as an explanation for the Seyfert core
emission. Although these rare, bright RSN, of the kind observed in
NGC~891 (van Gorkom et al. 1986), designated SN1986J and classified as
an unusual type II radio supernova (Weiler, Panagia \& Sramek, 1990),
can display high brightness temperatures at the peak of their light
curves (e.g. 10$^{9}$ K at $\lambda$6 cm at the peak of the SN1986J
light curve), their fluxes increase and decrease over a timescale of a
few years. The $\lambda$6-cm flux of SN1986J doubled in approximately
three years and halved again over the following three years. Measured
VLA fluxes for our Seyfert cores have typically varied by less than
30\% over $\sim$10 years and are therefore inconsistent with an RSN
interpretation. In addition, the {\em maximum} inferred 8.4-GHz
luminosity for SN1986J of $\sim$8~$\times$~10$^{20}$~W~Hz$^{-1}$
(calculated using the $\lambda$6-cm flux at the peak of the light
curve, $\alpha^6_{3.6}$~=~0.7, and assuming isotropic emission at a
distance of 8.96 Mpc) is somewhat lower than the luminosities of our
Seyfert nuclei (Table 2).

We therefore conclude that the high brightness temperatures ($>$
10$^8$ K), small sizes ($<$ 1 pc) and absence of strong flux
variations over $\sim$10-year timescales in T0109$-$383, NGC~2110,
NGC~5252 and Mrk~926 are probably indicative of synchrotron
self-absorption close to a jet-producing central engine. The lower
limits to the brightness temperatures do not require relativistic
motions, which is consistent with the non-relativistic proper motions
observed in two Seyfert galaxies by Ulvestad et al., (1999).

\subsection{Thermal Bremsstrahlung Emission in NGC~4388}

Unlike the other four galaxies discussed in this paper, no high
brightness temperature compact radio core is detected in NGC~4388 at
8.4 GHz. The observationally inferred brightness temperature of
2.4~$\times$~10$^4$~K~$<$~T$_{\rm B}$~$<$ 2.2~$\times$10$^6$~K (the
lower limit at 5 GHz, the upper limit at 8.4 GHz, see Section 4.3.1)
is too low for synchrotron self absorption to be important. Instead we
consider a model in which the emission is optically thin, thermal
bremsstrahlung, compatible with the observed flat radio
spectrum, from a gas with an electron temperature of T$_{\rm e}$~(K).

Our bremsstrahlung model consists of a thermal plasma of uniform
temperature, T$_{\rm e}$, and density, n$_e$, that fills the emitting
volume, $V$, with a filling factor $f$.  The maximum source volume is
set by the MERLIN beamsize and assumes an ellipsoidal geometry (with
semi-axes of 3.6~$\times$~1.5~$\times$~1.5 pc), while the minimum
source size, as discussed in Section 4.3.1, is $\sim$0.3~pc, for which
an edge-on disk geometry is assumed (disk diameter 0.4~pc, height
0.2~pc). Figures 4(a,b,c) show the allowed range of densities (n$_e$),
opacities at 4.993 GHz ($\tau_{\rm ff}$) and ionized gas column
densities (N$_{\rm e}$) as a function of T$_{\rm e}$, within the
limits set by the possible source sizes.

As shown by the shaded area in Figure 4(b), the gas becomes optically
thin at 4.993~GHz for T$_{\rm e}$~$>$~10$^{4.5}$~K, suggesting
n$_e$~$>$~1.6~$\times$~10$^4$~f~$^{-0.5}$~cm$^{-3}$ for the MERLIN
size limit, or T$_{\rm e}$~$>$~10$^{6.8}$~K and
n$_e$~$>$~1.8~$\times$~10$^6$~f~$^{-0.5}$~cm$^{-3}$ for the smaller
source size set by the VLBA limit. The larger electron density, for
$f$=1, is intermediate between that expected in Seyfert narrow line
regions (e.g., $\sim$10$^3$ cm$^{-3}$ - Koski, 1978) and broad line
regions ($\sim$10$^9$~cm$^{-3}$), consistent with the small size of
the region, and similar to ionized gas densities of
10$^5$--10$^7$~cm$^{-3}$ inferred from free-free absorption in the
inner parsec of Mrk~231 and Mrk~348 (Ulvestad et al., 1999). Similarly
high electron temperatures and densities have been derived for thermal
emission from the torus gas in NGC~1068 (Gallimore et al., 1997), thus
implying that we may be seeing the same phenomenon in NGC~4388.

The lower limit to the electron column density of N$_{\rm
e}$~$>$~7~$\times$~10$^{22}$~cm$^{-2}$ (Figure 4c) is compatible with
the total column of 4.2~$\times$~10$^{23}$~cm$^{-2}$ inferred from the
photoelectric absorption seen in the ASCA observed X-ray spectrum of
NGC~4388 (Iwasawa et al., 1997).  Demanding equality of ionized and
total columns would imply T$_{\rm e}$~$>$~10$^6$~K.

If the radio emission is indeed thermal bremsstrahlung, we may
calculate the predicted, intrinsic H$\beta$ flux from this gas using
(e.g., Ulvestad, Wilson \& Sramek, 1981):
$$\rm F(\rm H\beta)~({\rm erg~cm^{-2}~s^{-1}})~=~0.62~T_e^{0.5}({\rm
K})~g_{\rm ff}^{-1}({\rm \nu,~Z,~T})~\alpha^{\rm
eff}_{H\beta}~S_{\nu}({\rm mJy}),$$ where S$_{\nu}$ is the radio flux
density, $\alpha^{\rm eff}_{\rm H\beta}=3.01\times10^{-14}(\rm
T/10^4)^{-0.85}$ and g$_{\rm ff}$ is the Gaunt factor. Taking
S$_{\nu}$~=~2.1~mJy at 5 GHz, we predict
F(H$\beta$)~$<$~4.0~$\times$~10$^{-13}$~erg~cm$^{-2}$~s$^{-1}$ for
T$_{\rm e}$~$>$~10$^{4.5}$~K (Figure 4d). However the observed [O{\sc
iii}] flux of 1.86~$\times$~10$^{-13}$~erg~cm$^{-2}$~s$^{-1}$, through
1\farcs5$\times$3$''$ (Corbin et al., 1988) or 2$''\times$3\farcs5
apertures (Colina, 1992), and the [O{\sc iii}] to H$\beta$ ratio of
11.2 (Phillips \& Malin, 1982), result in a mean observed H$\beta$
flux of only 1.7~$\times$~10$^{-14}$~erg~cm$^{-2}$~s$^{-1}$ (which we
treat as an upper limit to the H$\beta$ flux from the MERLIN source
due to the large optical apertures compared to the radio sizes), a
factor of 24 lower than our predicted value if T$_{\rm
e}$~$=$~10$^{4.5}$~K.  If our assumption of a thermal origin for the
radio emission is correct, the difference between observed and
predicted F(H$\beta$) suggests A$_V$ $\gtrsim$ 3.0 magnitudes of
extinction towards the nucleus for T$_{\rm e}$=10$^{4.5}$~K. Using the
H$\beta$ flux measured by Dahari \& De Robertis (1988) of
5.2~$\times$~10$^{-14}$~erg~cm$^{-2}$~s$^{-1}$ results in a lower
extinction of A$_V$ $\gtrsim$ 1.9 mags.  In addition, no broad
Pa$\beta$, Br$\gamma$ or Br$\alpha$ lines are detected towards the
nucleus (Blanco et al., 1990; Ruiz et al., 1994; Veilleux et al,
1997), while a broad, off-nuclear H$\alpha$ line (Shields \&
Filippenko, 1988) is detected 4$''$ from the nucleus, consistent with
scattered emission from the broad line region. A temperature in excess
of 10$^8$~K would be required to explain the low observed H$\beta$
flux in the absence of extinction and therefore, given the implication
from optical studies that NGC~4388 harbors an obscured Seyfert type 1
nucleus, the moderately large value of extinction suggested by the
radio data does not seem unreasonable.

Alternatively, not all of the radio emission might be thermal, and the
base of a radio jet could provide a non-thermal contribution. However,
any significant non-thermal contribution would be difficult to
reconcile with the flat radio spectrum, given the low observed
brightness temperature. We therefore favor the thermal emission model
for the nuclear radio emission in NGC~4388.

\section{Conclusions}

We have used the VLBA at 8.4 GHz to study five Seyfert nuclei that
contain flat spectrum radio sources, in order to determine whether the
flat-spectrum nuclear radio emission, detected in VLA studies,
represents thermal emission from the accretion disk/obscuring torus or
synchrotron self-absorbed emission from a compact radio core source. 

Four of the five sources were detected (T0109$-$383, NGC~2110, NGC
5252, Mrk~926) and show compact, unresolved cores with brightness
temperatures, T$_{\rm B}$~$>$~10$^8$ K, total luminosities at 8.4 GHz
of $\sim$10$^{21}$ W Hz$^{-1}$ and sizes, on average, less than 1
pc. We conclude that the sub-pc scale radio emission in these sources
is non-thermal and self absorbed and, hence, dominated by the central
engine.  In addition to the core emission, NGC~2110 shows extended
emission which may represent the inner parts of the radio
jets. However, we find no evidence of thermal disk-like emission,
extended perpendicular to the collimation axis, in any of these
sources to a 3-$\sigma$ limit of $\sim$10$^6$ K.

The putative nucleus of NGC~4388 is not detected with the VLBA but is
detected with MERLIN at 5~GHz. The observationally inferred brightness
temperature of 2.4~$\times$~10$^4$~K~$<$~T$_{\rm
B}$~$<$~2.2~$\times$~10$^6$~K (the lower limit at 5~GHz, the upper
limit at 8.4~GHz) is too low for synchrotron self absorption to be
important.  Instead we have proposed a model in which the emission is
optically thin, free-free thermal bremsstrahlung emission from a gas
with an electron temperature of T$_{\rm e}$~$>$~10$^{4.5}$~K and
density n$_e$~$>$~1.6~$\times$~10$^4$~f~$^{-0.5}$~cm$^{-3}$ (where f
is the volume filling factor). The larger inferred values of T$_{\rm
e}$~$>$~10$^{6.8}$~K and
n$_e$~$>$~1.8~$\times$~10$^6$~f~$^{-0.5}$~cm$^{-3}$ for the smaller
source size set by the VLBA limit, are similar to the values of
$\sim$10$^{6.8}$~K and $\sim$10$^{6.8}$~cm$^{-3}$ found for thermal
emission from the torus gas in NGC~1068 (Gallimore et al., 1997), thus
implying that we may be seeing the same phenomenon in NGC~4388.
Sensitive VLBA observations of NGC~4388 at 1.4~GHz or 2.3~GHz are
required to spatially resolve the emitting region and determine its
exact physical properties.

\acknowledgments

We thank Pierre Ferruit, Neil Nagar and Dave Shone for useful
discussions, and Peter Thomasson for help with the MERLIN data. The
National Radio Astronomy Observatory is a facility of the National
Science Foundation operated under cooperative agreement by Associated
Universities, Inc. MERLIN is a U.K. national facility operated by the
University of Manchester on behalf of PPARC.  This research has made
use of: the United States Naval Observatory (USNO) Radio Reference
Frame Image Database (RRFID), NASA's Astrophysics Data System Abstract
Service (ADS) and the NASA/IPAC Extragalactic Database (NED), which is
operated by the Jet Propulsion Laboratory, California Institute of
Technology, under contract with the National Aeronautics and Space
Administration. This research was supported by the National Science
Foundation under grant AST 9527289 and NASA under grant NAG 81027.

\clearpage
\figcaption[]{Radio continuum images of the sources detected at 8.4
GHz with the VLBA - T0109$-$383, NGC~2110, NGC~5252 and Mrk~926. The
8.4-GHz VLBA images are shown in the right hand panels, with the
restoring FWHM beamsize shown as an ellipse in the bottom left corner
of each image and linear scale marked in the bottom right corner. A
VLA image of the larger scale structure of each source is shown in the
corresponding left hand panel (from Nagar et al. (1999) for NGC~2110
and NGC~5252, and from unpublished observations by J.A. Braatz,
L.L. Dressel \& A.S. Wilson for T0109$-$383 and Mrk~926). Contour levels
for the VLBA images and unpublished VLA images are given in Table 5.}

\vspace*{5mm}

\figcaption[]{MERLIN 5-GHz radio continuum image of NGC~4388. The
nucleus is marked M1 and the secondary weak feature (possibly a
component in the radio jet) is marked M2. The beamsize is indicated by
an ellipse in the lower left corner. The contour levels, in multiples
of 3$\times$r.m.s., are (--2, --1, 1, 2, 3, 4) $\times$ 0.3 mJy beam$^{-1}$
or T$_{\rm B}$ = (--2, --1, 1, 2, 3, 4) $\times$ 5.9$\times$10$^3$~K
(beamsize 91.0 $\times$ 39.5 mas).}

\vspace*{5mm}

\figcaption[]{(a) VLA 8.4-GHz radio continuum image (full resolution
image produced from data presented by Falcke et al., 1998) of the
1\farcs9 central double source in NGC~4388 (each component is marked
V1 and V2, with V1 taken to be the core); (b) Super-resolved VLA image,
from the same $(u,v)$ data as used for (a), revealing collimated
ejection from the core and bending of the jet at V2. The crosses
indicate the two MERLIN components (M1 and M2), where M1 is coincident
with the core and M2 coincides with a protrusion in the 8.4-GHz VLA
contours in the direction of the jet. The contour levels for both
images are (--1, 1, 2, 4, 8, 16, 32) $\times$ 90 $\mu$Jy beam$^{-1}$
and the beamsizes (shown in the lower left corner of each image) are
(a) 0\farcs70 $\times$ 0\farcs22 and (b) 0\farcs22 $\times$
0\farcs22.}

\vspace*{5mm}

\figcaption[]{Range of possible values of (a) electron density, n$_e$,
(b) free-free opacity at 4.993 GHz, $\tau_{\rm ff}$, (c) column
density of ionized gas, N$_{\rm e}$, and (d) predicted H$_{\beta}$
flux, assuming the radio emission from NGC~4388, at 4.993 GHz, is
free-free thermal bremsstrahlung emission from an ionized gas with an
electron temperature, T$_{\rm e}$. Upper and lower limits to the
plotted quantities are derived from the lower and upper limits to the
source size (Section 4.3.1); we assumed the maximum source volume
(MERLIN limit) corresponds to an ellipsoid (with semi-axes
3.6~$\times$~1.5~$\times$~1.5~pc), and the minimum volume (VLBA limit)
corresponds to an edge-on disk (with diameter, 0.4~pc, and height,
0.2~pc). The permitted values for an optically thin plasma (as
indicated by the flat radio spectrum) are indicated by the shading.}

\clearpage

\setcounter{table}{0}
\begin{table}
\caption{Observing parameters for the 8.4-GHz VLBA observations of
Seyfert galaxies and calibrators. Measured positions, derived from
Gaussian fitting, for the check sources are also given. The surveys
from which the positions of the phase calibrators were
selected are listed below as footnotes, along with the corresponding
positional accuracy.}
\scriptsize
\begin{tabular}{lccccc} \\
\tableline
Target Seyfert &T0109$-$383& NGC~2110 & NGC~4388 & NGC~5252 & Mkn~926 \\
\tableline\\
Date of observations &07 Aug 1997 &9 Aug 1997 &12 Jun 1998 & 14 Jun 1998&21 Jul 1997  \\
\\       
Phase calibrator&J0106$-$4034$^a$& J0541$-$0541$^b$ & J1207+1211$^c$& J1320+0140$^d$ & J2255$-$0844$^a$\\
Position used (J2000)&01h06m45.1080s& 05h41m38.0834s
&12h07m12.625s&13h20m26.7938s&22h55m04.2398s  \\
		    &--40$^{\circ}$34$'$19\farcs960 & --05$^{\circ}$41$'$49\farcs428 &12$^{\circ}$11$'$45\farcs89&01$^{\circ}$40$'$36\farcs786 & --08$^{\circ}$44$'$04\farcs022\\
Source + phase cali- && \\
 -brator cycle time&$\sim$2 + 1 min&$\sim$4 + 1 min &$\sim$3 + 1 min&$\sim$3 + 1 min &$\sim$4 + 1 min \\
\\
`Check' calibrator&J0044$-$3530& J0607$-$0834& J1214+0829&J1359+0159&J2246$-$1206\\
Position used (J2000)&00h44m41.229s &06h07m59.699s
&12h14m59.914s&13h59m27.1478s& 22h46m18.232s \\
		     &--35$^{\circ}$30$'$41\farcs63&--08$^{\circ}$34$'$49\farcs98&08$^{\circ}$ 29$'$
22\farcs53&01$^{\circ}$ 59$'$ 54\farcs543&--12$^{\circ}$06$'$51\farcs28 \\
Position measured & --- &06h07m59.6975s&12h14m59.913s&13h59m27.1512s&22h46m18.2313s\\
	&	&--08$^{\circ}$34$'$49\farcs994&08$^{\circ}$29$'$22\farcs56 &01$^{\circ}$59$'$54\farcs531&--12$^{\circ}$06$'$51\farcs262\\
Times on `check' cal.&3 $\times$ 1 min&4 $\times$ 1 min&12 $\times$ 1 min&13 $\times$ 1 min&4 $\times$ 1 min\\
\\
Online fringe-finder& 3C454.3& DA193 & 3C273 & 3C273& 3C454.3\\
\\
Total time on source&$\sim$2.5 hrs&$\sim$3.2 hrs&$\sim$6.0 hrs&$\sim$5.8 hrs&$\sim$3.3 hrs\\
\tableline 
\multicolumn{6}{l}{$^{a,b}$Ma et al., 1998 (2.5, 0.4 mas) $^c$Browne et al., 1998;Wrobel et al., 1998 (14 mas) $^d$Peck \& Beasley, 1998 ($<$5 mas)}
\end{tabular}
\end{table}

\begin{table} 
\caption{Results of VLBA observations. Galaxy names, positions, peak
and integrated fluxes from single component Gaussian fitting, and
beamsizes ($\theta$) are listed (columns 1--5). Lower limits to the
brightness temperatures (T$_B$ (K), column 6) are derived from the
given peak flux, assuming an upper limit to the source size of half of
the beamsize; for NGC~2110, the peak flux of the unresolved component
(given in Table 3) is used instead. The luminosity at 8.4 GHz (L,
column 7) is calculated using the total flux density and assuming
isotropic emission at the distance of the galaxy (D, column 8; obtained
assuming H$_0$ = 75 km s$^{-1}$ Mpc$^{-1}$ and q$_0$ = 0.5).}
\scriptsize
\begin{tabular}{lcccccccc} \\
\tableline
&Fitted position &8.4-GHz&8.4-GHz&$\theta$
    &T$_B$ (K)&L& D\\ 
  
Seyfert&(J2000.0)(R.A.)&Peak Flux& Total Flux& (mas)&($\times$10$^8$ K)&($\times$
10$^{21}$)&(Mpc)\\
       &        (J2000.0)(Dec.)&(mJy beam$^{-1}$) & (mJy)  &({\em pc})&& (W Hz$^{-1}$) \\
\tableline
T0109$-$383 &  01 11 27.6413 & 9.9 $\pm$0.6 &
11.0$\pm$  0.7& 1.96 $\times$ 0.63&$>$8.1&2.7&46.6\\
      &        --38 05 00.477  &&&{\em 0.45 $\times$ 0.14}\\
NGC~2110$^a$ & 05 52 11.3762  &  16.6 $\pm$ 0.8&
29.5$\pm$1.5  & 1.88 $\times$ 0.72&$>$6.0&3.1&30.4\\
      &        --07 27 22.513    &&&{\em 0.28 $\times$ 0.11}\\
NGC~5252 &  13 38 15.8698 & 7.9 $\pm$0.4 &
9.1$\pm$  0.5& 2.01 $\times$ 0.95&$>$4.2& 8.8&92.4 \\
      &     +04 32 33.513  &&&{\em 0.91 $\times$ 0.43}\\
Mkn~926 &  23 04 43.4776 & 4.6 $\pm$0.2 &
5.0$\pm$  0.3& 2.82 $\times$ 1.02&$>$1.7&2.1&191.4\\
      &        --08 41 08.629  &&&{\em 2.64 $\times$ 0.95}\\
\tableline
\end{tabular}

$^a$See Table 3 for properties of multiple
components present in NGC~2110.
\end{table}

\newpage
\begin{table} 
\centering 
\caption{Results of Gaussian fitting to the emission from NGC~2110;
three components are detected, which are identified (column 1) as an
unresolved `core', north-south `jet' emission and a northern
`knot'. The positions of each component (column 5) relative to the core
position (column 2) are given with positive offsets north of the
core. Peak and integrated fluxes are given in columns 3 \& 4, along with
deconvolved sizes and position angles (column 6, 7) for the marginally
resolved components.}

\scriptsize
\begin{tabular}{lcccccc} \\

\tableline

&Fitted position &8.4 GHz &8.4-GHz&Offset of
    fitted  &Deconvolved size & PA\\
  
Component&(J2000.0)(R.A.)&Peak Flux& Total Flux& component centroid,
         & (mas)& (deg)\\
       &(J2000.0)(Dec.)&(mJy beam$^{-1}$)&(mJy)&$\Delta$mas relative to `core'    &\\
\tableline
Unresolved `core' & 05 52 11.3762 & 8.0 $\pm$ 0.5&7.5 $\pm$0.4 & 0
      & 0&0\\ 
& --07 27 22.513 &&&\\
N-S `jet' & - &8.7$\pm$0.5&13.1 $\pm$ 0.7  &
      +0.21&1.41 $\times$ 0.36&8.5\\
      &        &&&&($\pm$0.3) ($\pm$0.3)&($\pm$0.8)\\

Northern `knot'&- &5.0 $\pm$ 0.3 &6.0 $\pm$ 0.3& +1.95 &
0.75 $\times$ 0.12 &137\\
&&&&&($\pm$0.4) ($\pm$0.12)&($\pm$90)\\
\tableline
\end{tabular}
\end{table}

\begin{table} 
\centering 
\caption{Upper table (a) shows the results of Gaussian fitting to the
5-GHz core emission in NGC4388, imaged with MERLIN. Fitted position,
peak brightness, integrated flux density and beam size are given. The
lower table (b) shows the corresponding derived brightness temperature
T$_B$(K) (calculated using the measured MERLIN peak brightness and
beamsize) and the {\em predicted} VLBA brightness temperatures at 8.4
GHz, assuming the spectral indices given, the VLBA beamsize given and
that the source is unresolved with the VLBA. These predicted VLBA
brightness temperatures exceed the observed value, allowing a lower
limit to the source size to be determined (Section 4.3.1).}

\scriptsize
\begin{tabular}{lcccccccc}\\ 
\tableline
\tableline
\multicolumn{7}{c}{\bf (a) Results of Gaussian fitting to 5 GHz MERLIN
image of NGC4388.}\\
\tableline
Seyfert&Fitted position &5-GHz Peak Flux&5-GHz Integrated Flux&MERLIN Beamsize\\
       &(B1950.0)(R.A.)&(mJy/beam)&(mJy)&(mas)\\
       &(B1950.0)(Dec.)& \\
\tableline
NGC4388 & 12 23 14.646 &  1.2$\pm$0.1 &2.1$\pm$0.3&91.0~ $\times$
~39.5\\
      &     12 56 20.20     &&&&&\\
\tableline
\tableline
\multicolumn{7}{c}{\bf (b) Derived MERLIN 5-GHz  and
 predicted VLBA 8.4-GHz brightness temperatures.}\\
\tableline 
&Derived T$_B$ (10$^4$ K)&VLBA Beamsize&Assumed $\alpha$ from&{\em Predicted} T$_B$ (10$^6$ K)\\
&(at 5 GHz)& & $\nu$=5 to  8.4 GHz & (at 8.4 GHz)\\
\tableline
&$>$2.4&2.52$\times$1.46 mas& 0  &8.3\\
&	&(0.23$\times$0.14 pc)       &0.26&7.3 \\

\tableline
\tableline
\end{tabular}
\end{table}

\begin{table}
\centering
\caption{Upper table lists contour levels (column 2), plotted as
multiples of 3~$\times$~r.m.s in the image (column 3), and beamsizes
(column 4) for the VLBA images in Figure 1. Brightnesses (mJy
beam$^{-1}$) are converted to T$_{\rm B}$~(10$^6$~K) using the
corresponding VLBA beamsizes. Lower table lists contour levels (column 2)
for the unpublished VLA images of T0109$-$383 and Mrk~926 shown in
Figure 1, plotted as multiples of 3~$\times$~r.m.s in the image
(column 3). The VLA beamsizes are given in column 4.}

\scriptsize
\begin{tabular}{lccccc} \\
\tableline
Seyfert & 8.4-GHz VLBA Contour Levels& 3~$\times$~r.m.s.&Beamsize\\
 &	(multiples of 3 $\times$ r.m.s.)&(mJy beam$^{-1}$)~(10$^6$ K)&(mas)
	\\
\tableline
T0109$-$383&   --1, 1, 2, 4, 8, 16 &~0.45~~~~~~~~~~~~9.2&1.96 $\times$ 0.63\\
NGC~2110 &  --1, 1, 2, 4, 8, 16, 32 &~0.28~~~~~~~~~~~~5.3& 1.88 $\times$ 0.72\\
NGC~5252 & --1, 1, 2, 4, 8, 16 &~0.30~~~~~~~~~~~~4.0 &2.01 $\times$ 0.95 \\
Mkn~926  &  --1, 1, 2, 4, 8, 16 &~~0.14~~~~~~~~~~~~1.2&2.82 $\times$ 1.02 \\
\tableline 
 & 8.4-GHz VLA Contour Levels& 3~$\times$~r.m.s.&Beamsize\\ 
	&(multiples of 3 $\times$ r.m.s.)&(mJy beam$^{-1}$)& ($''$)\\
\tableline 
T0109$-$383&   --1, 1, 2, 4, 8, 16, 32, 64 &0.11&0.41 $\times$ 0.17\\
Mkn~926  &  --1, 1, 2, 4, 8, 16, 32 &0.12& 0.27 $\times$ 0.21 \\
\tableline
\end{tabular}
\end{table}
\clearpage

{\bf Figure 1}
\begin{figure}
\plotfiddle{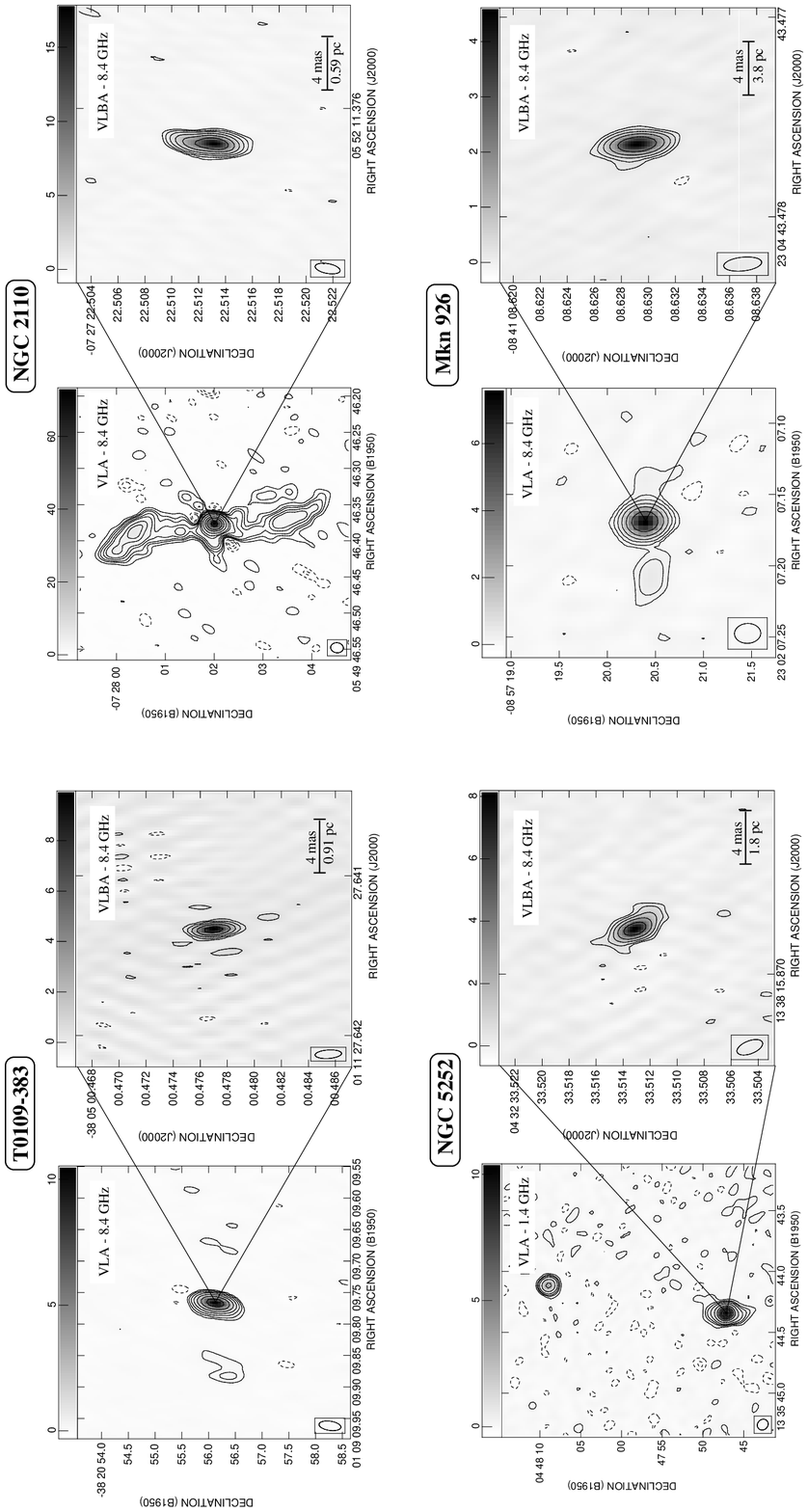}{1cm}{0}{92}{92}{-230}{-90}

\end{figure}
\clearpage

{\bf Figure 2}
\begin{figure}
\plotfiddle{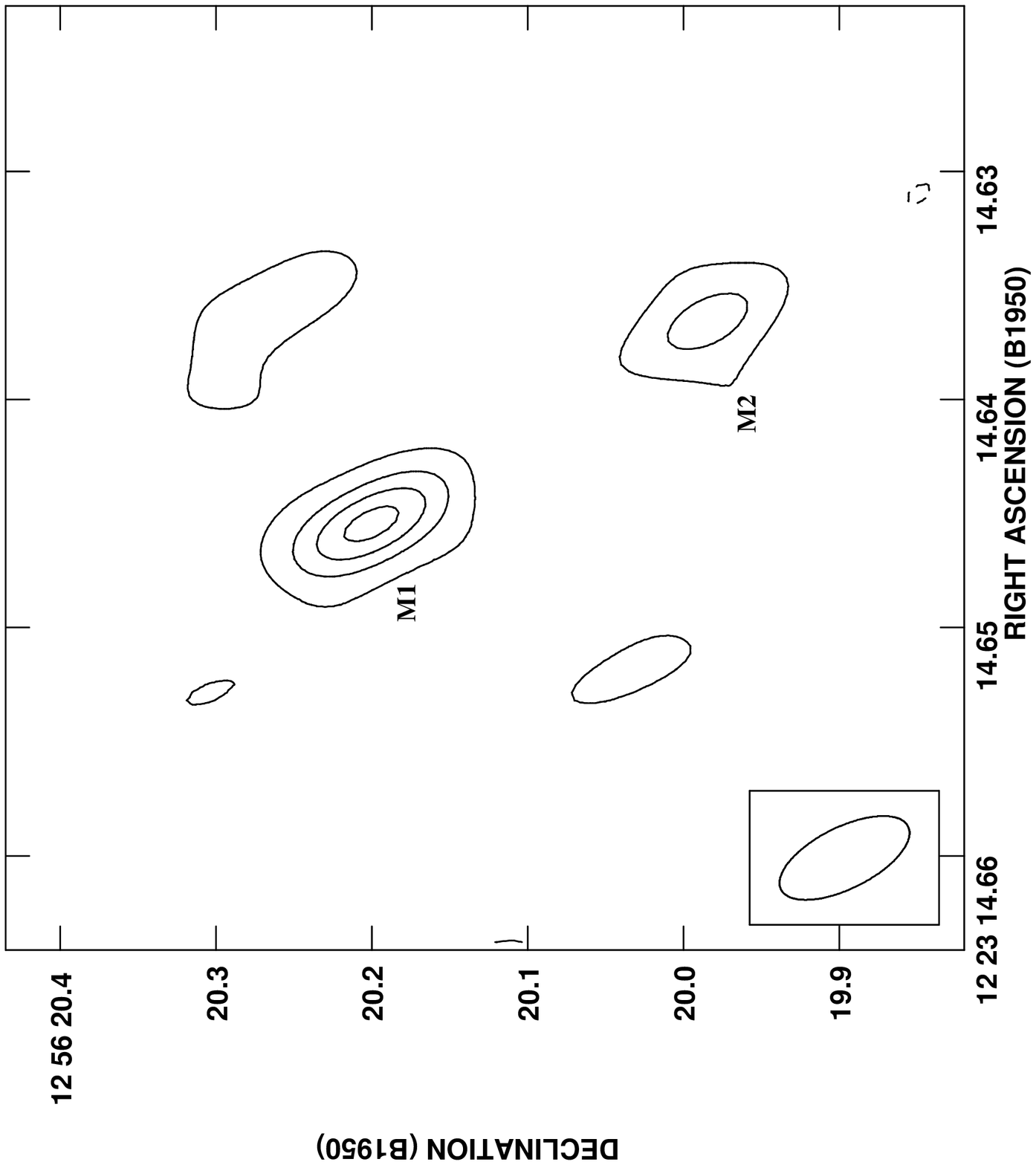}{1cm}{270}{80}{80}{-300}{550} 
\end{figure}
\clearpage

{\bf Figure 3}
\begin{figure}
\plotfiddle{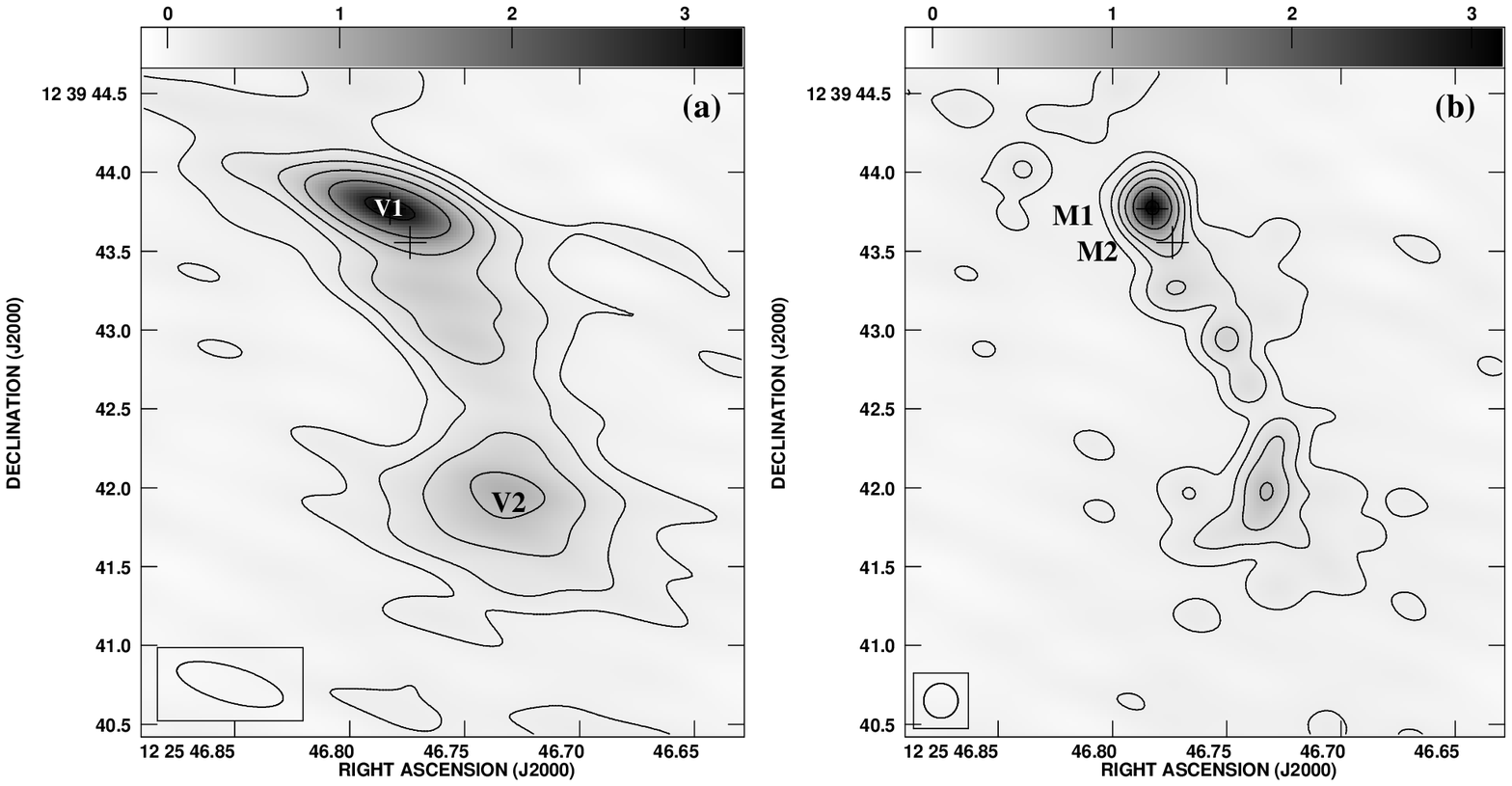}{1cm}{90}{100}{100}{200}{-30}
\end{figure}
\clearpage

{\bf Figure 4}
\begin{figure}
\plotfiddle{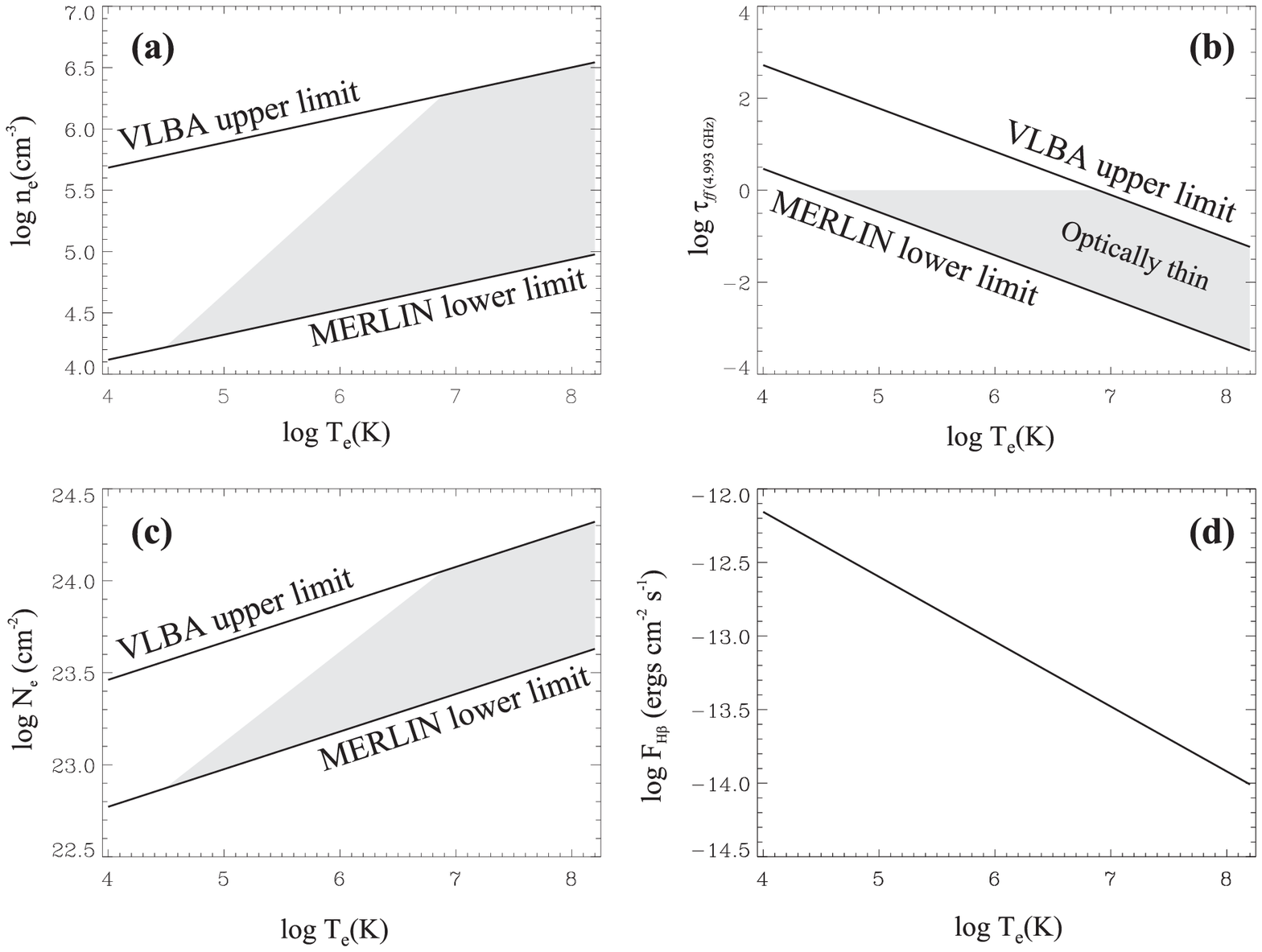}{1cm}{90}{110}{110}{630}{-80}
\end{figure}

\end{document}